\documentclass[prl ,aps, english, twocolumn, hyperref]{revtex4}

\usepackage{graphicx}

\usepackage{amssymb, amsmath, color, rotating, multirow}

\begin{document}

\title{Pairing, Ferromagnetism, and Condensation of a normal spin-$1$ Bose gas}

\author{Stefan S. Natu}

\email{ssn8@cornell.edu}

\affiliation{Laboratory of Atomic and Solid State Physics, Cornell University, Ithaca, New York 14853, USA.}

\author{Erich J. Mueller}

\affiliation{Laboratory of Atomic and Solid State Physics, Cornell University, Ithaca, New York 14853, USA.}

\begin{abstract}
We theoretically study the stability of a  normal, spin disordered, homogenous spin-$1$ Bose gas against ferromagnetism, pairing, and condensation through a Random Phase Approximation which includes exchange (RPA-X). Repulsive spin-independent interactions stabilize the normal state against both ferromagnetism and pairing, and for typical interaction strengths leads to a direct transition from an unordered normal state to a fully ordered single particle condensate. Atoms with much larger spin-dependent interaction may experience a transition to a ferromagnetic normal state or a paired superfluid, but, within the RPA-X, there is no instability towards a normal state with spontaneous nematic order. We analyze the role of the quadratic Zeeman effect and finite system size.
\end{abstract}
\maketitle

\textit{Introduction.---}
The interplay of superconductivity/superfluidity and magnetism is fundamental \cite{scpapers}. Experiments in ultra-cold spin-$1$ gases \cite{stamper, mukund, sadler, sengstock} have begun to explore this physics, elucidating the subtle connections between Bose condensation of single particles and competing/complementary orders such as pair condensation, ferromagnetism, and liquid-crystal like nematicity \cite{stenger,ho,ohmi}.  This has become a model system for thinking about exotic spin textures and topological defects  \cite{erich, mukherjee1, podolsky, zhou2},  and the dynamics of quantum phase transitions \cite{lamacraft, ueda}. However, the finite temperature $3$D phase diagram is a mystery, with earlier works producing contradictory results \cite{yang, gu, Szabo}.  Here we clarify the situation by using a well-controlled approximation [the random phase approximation with exchange (RPA-X)] to calculate the instabilities of the normal state.   

A particularly dramatic feature of the spin-1 Bose gas is that it supports a bosonic analog of the BCS transition \cite{evans,Nozieres,bardeen, erich1}. Somewhat counterintuitively, the paired state is {\em less ordered} than a single particle condensate, and is found when both the spin independent and dependent interactions are repulsive.  In addition to its theoretical importance, this feature makes it an interesting paradigm to keep in mind while exploring the mechanisms for superconductivity in systems such as high-Tc cuprates, C-60 and polyacenes where the interactions are believed to be repulsive \cite{chakravarty}. 

The Hamiltonian of a spin-$1$ Bose gas is the sum of a kinetic and interaction term,
$H={\hat{\cal{H}}}_{kin}+ {\hat{\cal{H}}}_{int}$.  In the presence of a magnetic field in the $\hat z$ direction, the kinetic term has the form
${\hat{\cal{H}}}_{kin}=\sum_{k\sigma} \epsilon_{k\sigma} a_{k\sigma}^\dagger a_{k\sigma}$, where $a_{k\sigma}$ is the annihilation operator for a boson with
momentum $k$ and spin projection $\sigma=-1,0,1$.  The dispersion is $\epsilon_{k0}=k^2/2m-\mu$, $\epsilon_{k\pm1}=k^2/2m-\mu+q\pm p$, where $p$/$q$ are linear/quadratic in the magnetic field.  There is no spin-orbit coupling, allowing us to eliminate the linear Zeeman effect ($p$) by working in a rotating frame.  Off-resonant microwave light allows the quadratic Zeeman field $q$ to be tuned, taking on positive and negative values \cite{gerbier}.  Assuming short range interactions, symmetry forces the interaction Hamiltonian to be \cite{ho, ohmi}:
\begin{equation}\label{eq:1}
{\hat{\cal{H}}}_{int} = \frac{1}{2}\int d\textbf{r}~\psi^{\dagger}_{\alpha}\psi_{\beta}^{\dagger}\psi_{\gamma}\psi_{\delta}(c_{0}\delta_{\alpha\delta}\delta_{\beta\gamma} + c_{2}\hat{\textbf{S}}_{\alpha\delta}\cdotp \hat{\textbf{S}}_{\beta\gamma}), 
\end{equation}
where the greek indices denote the spin projection and $\psi_{\alpha}(r)=\frac{1}{V}\sum_k e^{ik r} a_{k\sigma}$ is the the boson field operator. 

The two coupling constants, $c_0$ and $c_2$ represent spin independent and spin dependent interactions  
The $\hat{\textbf{S}}$ operators denote $3\times3$  spin-$1$ matrices. The interactions are expressed in terms of the microscopic scattering lengths in the spin-$0$ ($a_{0}$) and spin-$2$ ($a_{2}$) channels and atomic mass $m$ as: $c_{0} = 4\pi (a_{0}+ 2a_{2})/3m$ and $c_{2} = 4\pi(a_{2} -a_{0})/3m$. 

Two atoms are typically used in these experiments: $^{87}$Rb ($c_2 < 0$) and $^{23}$Na ($c_2 > 0$). In all experiments so far, spin independent interactions are repulsive ($c_{0} > 0)$ and $c_{0} \gg |c_{2}|$. We find that the phase diagram is featureless in this regime, motivating us to study the more general case where the interactions are comparable in magnitude. Perhaps $^7$Li, rare earth atoms, or alkali-earth atoms will have scattering parameters in this regime.
 Although
dipolar interactions are believed to play  an important role in the low temperature quasi-$2$D experiments of  Vengalattore \textit{et al.} \cite{mukund}, we neglect them here, as they 
are  much too weak to influence the stability of the normal state.

\begin{table}[tb]
\caption{Orders in spin-1 gas.}\label{tab}
\begin{ruledtabular}
\begin{tabular}{lcr}
\bf Order&\bf Symbol&\bf Order Parameter\\
ferromagnetic&\bf F&$\langle S \rangle \neq 0$\\
nematic&\bf N&$\langle S_{\mu}S_{\nu} \rangle \neq \delta_{\mu\nu}$\\
single particle&\bf C& $\langle\psi_{\mu}\rangle \neq 0$\\
pair&\bf P&$\sum_{k}\langle a^{\dagger}_{\mu \textbf{k}}a^{\dagger}_{\nu \textbf{-k}}\rangle\neq 0$
\end{tabular}
\end{ruledtabular}
\end{table}

The spin-1 gas can present several types of order, summarized in Table~\ref{tab}. Single particle condensate order ({\bf C}) is always accompanied by either ferromagnetic ({\bf FC}) or nematic order ({\bf NC}).  Mean field examples of these condensed  states are $|FC\rangle= (\psi^{\dagger}_{1})^{N}|0\rangle$, and $|NC\rangle= (\psi^{\dagger}_{0})^{N}|0\rangle$, with the former seen in $^{87}$Rb, and the latter in $^{23}$Na.  The singlet state  from \cite{law} with all particles in $k=0$, 
$|S\rangle=
  ((\psi^{\dagger}_{0,{\bf k=0}})^{2}
   -2 \psi^{\dagger}_{1,{\bf k=0}}
    \psi^{\dagger}_{-1,{\bf k=0}}
     )^{N/2}|0\rangle$, has off-diagonal single-particle order [ie. $\lim_{|r-r^\prime|\to\infty} \langle \psi_0^\dagger(r)\psi_0(r^\prime)\rangle\neq0$] and in the thermodynamic limit should be considered as an {\bf NC} state \cite{erich2}.    An example of a {\bf P} state would be a condensate of small singlet pairs
$|P\rangle=\kappa^{N/2}|0\rangle$, where $\kappa= \sum_{k}\left(a^{\dagger}_{0 \textbf{k}}a^{\dagger}_{0 \textbf{-k}}- 2 a^{\dagger}_{1  \textbf{k}} a^{\dagger}_{-1 -\textbf{k}}\right)$.  Unlike $|S\rangle$, the state $|P\rangle$ has no off-diagonal single particle order.  Paired states for which
$\langle a^{\dagger}_{0 \textbf{k}}a^{\dagger}_{0 \textbf{-k}}- 2 a^{\dagger}_{1  \textbf{k}} a^{\dagger}_{-1 -\textbf{k}} \rangle\neq0$ posses nematic order.  We find no instabilities towards paired states with ferromagnetic order.

\begin{figure}
\begin{picture}(100, 90)
\put(-20, -7){\includegraphics[scale=0.5]{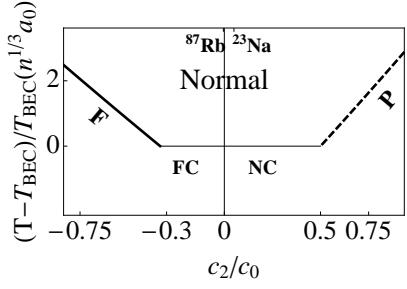}}
\end{picture}

\caption{\label{fig:-1}\textbf{3D Phase diagram at $\bf{q=0}$ within the RPA-X}: Thick solid/dashed lines are the ferromagnetic and pairing transition temperatures measured from the ideal Bose gas transition $k_{B} T_{BEC} = \frac{2\pi}{m}(\frac{n}{3\zeta(3/2)})^{2/3}$ and scaled by $n^{1/3} a_0$ as a function of  spin-dependent interaction $c_{2}$. Thin solid line shows the instability towards single particle condensation (ferromagnetic or polar depending on the sign of $c_{2}$).
For $c_{2}>0.5c_{0}$, the normal state is unstable to a rotationally symmetric paired singlet phase (P) with a $T_{c} > T_{BEC}$. For $c_{2} < -c_{0}/3$, the normal state becomes unstable to a  ferromagnet (F). For $|c_{2}|$ smaller than these threshold values, there is a direct transition from a disordered normal state to a ferromagnetic condensate (FC) or a nematic condensate (NC). Tick marks on the upper frame illustrate the scattering lengths for $^{85}$Rb and $^{23}$Na.}
\end{figure}

Although the 2D phase diagram is well established (with an algebraically ordered $\bf P$ state at any finite temperature when $q=0$) \cite{mukherjee1,yang},
contradictory results have appeared concerning the $3$D phase diagram. Both Gu and Klemm \cite{gu} and Kun Yang \cite{yang} erroneously found that arbitrarily weak \textit{attractive} interactions drive a ferromagnetic instability with $T^{F}_{c} > T_{BEC}$. Kis-Szab\'{o}, Sz\'{e}falusy and Szirmai \cite{Szabo} gave a more thorough argument, finding a finite threshold for this instability. We extend their calculation to incorporate exchange physics, including $c_{2} > 0$, $c_{2} < 0$, and the quadratic Zeeman effect. 

\textit{Formalism.---} We calculate the longitudinal ($\chi_{z}$) and transverse spin ($\chi_{\pm}$) and pairing ($\Pi$) susceptibilities of the homogeneous interacting spinor Bose gas using a Hartree-Fock Random Phase Approximation (RPA-X). A divergence of the zero frequency, long wavelength susceptibility, $\chi^{-1}(\textbf{k}=0 , \omega=0) = 0$ signals an instability in that channel.  Within the RPA-X there is never an instability in the nematic channel which is not simultaneously accompanied by single particle or pair condensation.  

The relevant response functions are
\begin{eqnarray} \label{eq:2}
\chi^{\gamma\delta}_{\alpha\beta \textbf{p}} (t) = \frac{1}{i}\langle \sum_{k, q}a^{\dagger}_{\delta\textbf{k}}(t)a_{\gamma\textbf{k}+ \textbf{p}}(t)a^{\dagger}_{\beta \textbf{q}}(0)a_{\alpha \textbf{q}- \textbf{p}}(0)\rangle\\\label{eq:9}
\Pi^{\gamma\delta}_{\alpha\beta \textbf{p}} = \frac{1}{i} \langle \sum_{k, q}a^{\dagger}_{\delta \textbf{k}}(t)a^{\dagger}_{\gamma \textbf{p}-\textbf{k}}(t)a_{\beta \textbf{q}}(0)a_{\alpha \textbf{p}-\textbf{q}}(0)\rangle
\end{eqnarray}
where $t>0$, and the greek subscripts denote spin indices and $\textbf{p}$ is the momentum \cite{Negele, Szirmai}. The longitudinal and transverse spin correlation functions are $\chi_{z\textbf{p}}(t) = -i\langle S_{z \textbf{p}}(t) S_{z -\textbf{p}}(0) \rangle = -\sum_{\alpha,\delta = \pm 1}(-1)^{\alpha\delta}\chi^{\delta\delta}_{\alpha\alpha \textbf{p}}(t)$, and
$\chi_{\pm}=-i\langle S_{+ \textbf{p}}(t) S_{- -\textbf{p}}(0) \rangle =\chi^{10}_{01}+\chi^{0\phantom{-}1}_{0-1}+\chi^{-1\phantom{-}0}_{\phantom{-}0-1}+\chi^{-10}_{\phantom{-}10}$, where $S_{\mu \textbf{p}}(t) = \sum_{\textbf{q}}a^{\dagger}_{\textbf{p}+ \textbf{q}/2}(t) \hat S_{\mu}a_{\textbf{p}- \textbf{q}/2}(t)$, and $\mu = \{z, \pm \}$.

In the RPA-X, the susceptibility of the interacting gas is determined from the non-interacting susceptibility by summing over all repeated direct, and exchange interactions. 
\begin{eqnarray} \label{eq:3}
(\chi^{RPA})^{ \gamma\eta}_{\alpha\beta} = (\chi^{0})^{ \gamma\eta}_{\alpha\beta}\delta_{\alpha\eta}\delta_{\beta\gamma} + \sum_{\mu\nu}(\chi^{0})^{ \gamma\eta}_{\eta\gamma} \textbf{V}^{\gamma\eta}_{\mu\nu}(\chi^{RPA})^{ \nu\mu}_{\alpha\beta} \\\label{eq:10}
(\Pi^{RPA})^{\gamma\eta}_{\alpha\beta} = (\Pi^{0})^{\gamma\eta}_{\alpha\beta}\delta_{\alpha\eta}\delta_{\beta\gamma} + \sum_{\mu\nu}(\Pi^{0})^{ \gamma\eta}_{\eta\gamma} \textbf{V}^{\gamma\eta}_{\mu\nu}(\Pi^{RPA})^{ \mu\nu}_{\gamma\eta} 
\end{eqnarray}
The interaction potential $V_{\mu\nu}^{\gamma\eta}$ of Eq.~(\ref{eq:1}), which includes both direct and exchange graphs, is explicitly given in the supplementary material. 

The non-interacting Green's functions are diagonal in spin space: $(\chi^{0})^{\gamma\eta}_{\alpha\beta \textbf{p}} (t)  = 0$, and $(\Pi^{0})^{\gamma\eta}_{\alpha\beta \textbf{p}} (t)  = 0$ unless $\eta = \alpha$ and $\gamma = \beta$,
\begin{eqnarray}\label{eq:4}
(\chi^{0})^{\beta\alpha}_{\alpha\beta}(p, \omega) &=& \int \frac{d^{3}\textbf{k}}{(2\pi)^{3}}\frac{n(\epsilon_{k, \alpha}) - n(\epsilon_{k+p, \beta})}{\omega - (\epsilon_{k+p \beta} - \epsilon_{k\alpha})}\\\label{eq:11}
(\Pi^{0})^{\alpha\beta}_{\beta\alpha}(\textbf{p}, \omega) &=& \int \frac{d^{3}\textbf{k}}{(2\pi)^{3}}\frac{n(\epsilon_{\textbf{k}, \alpha}) + n(\epsilon_{\textbf{k}+\textbf{p}, \beta})}{\omega - (\epsilon_{\textbf{k}+\textbf{p} \beta} + \epsilon_{\textbf{k}\alpha})}
\end{eqnarray}
Here $n(\epsilon_{k \sigma})=(e^{\beta \epsilon_{k\sigma}}-1)^{-1}$ is the Bose-Einstein distribution at temperature $T=1/\beta$. 
For a non-interacting gas, the spin susceptibility $\chi^{0}$, pairing susceptibility $\Pi^0$ and compressibility all diverge as $\mu \rightarrow 0$ from below, marking Bose-Einstein condensation. 

At $k,\omega=0$, these non-interacting response functions may be written in terms of the polylogarithm functions $g_\nu(z)=\sum_j z^j/j^{\nu}$: 
$\chi^{\phantom{-}1-1}_{-1\phantom{-}1}(\textbf{k}=0, \omega=0) =  - \frac{m}{2\pi\Lambda_{T}} g_{1/2}(e^{\beta(\mu - q)})$, 
$(\chi^{0})^{10}_{01}(0, 0) = \frac{m}{4\pi\Lambda_{T}}(T/q)[g_{3/2}(e^{\beta(\mu - q)}) - g_{3/2}(e^{\beta\mu})]$, $\Pi^{\beta\alpha}_{\alpha\beta}(0, 0) = -\frac{m}{\pi\Lambda_{T}}g_{1/2}(e^{\beta\mu_{\text{eff}}})$, where $\mu_{\text{eff}}  = \mu - q$ for $\alpha = \pm1$ and $\beta = \mp 1$, and $\mu_{\text{eff}}  = \mu$ for $\alpha = \beta = 0$. The thermal wavelength is $\Lambda_T=\sqrt{2\pi/m k_B T}$. The calculations are detailed in the supplementary material.

To detect ferromagnetism we consider the response functions (see supplementary materials)
\begin{eqnarray}\label{eq:RPA}
\chi^{RPA}_{z}(\textbf{k}, 0) &=& \frac{2(\chi^{0})^{\phantom{-}1-1}_{-1\phantom{-}1}(\textbf{k}, \omega)}{1 - (c_0 + 3 c_{2})(\chi^{0})^{\phantom{-}1-1}_{-1\phantom{-}1}(\textbf{k}, 0)}\\
\chi^{RPA}_{\pm}(\textbf{k}, 0) &=& \frac{2(\chi^{0})^{10}_{10}(\textbf{k}, 0)}{1 - (c_0 + 3 c_{2})(\chi^{0})^{10}_{01}(\textbf{k}, 0)}
\end{eqnarray}
To detect pairing, it suffices to consider the singlet pairing susceptibility, $\Theta  = (\Pi^{00}_{00} -2 \Pi^{\phantom{-}1-1}_{-1\phantom{-}1})^{RPA}$,
\begin{eqnarray}\nonumber
\Theta &=&\frac{\Pi^{+} - 2\Pi^{0}+ \Pi^{0}\Pi^{+}(c_0 - 2c_2)}{1 - (c_{0}-c_{2})\Pi^{+} - c_{0}\Pi^{0} + (c_{0} - c_{2})(c_{0}+c_{2})\Pi^{+}\Pi^{0}}.
\end{eqnarray}
with $\Pi^{0} \equiv  (\Pi^{0})^{00}_{00}$ and $ \Pi^{+} \equiv  (\Pi^{0})^{\phantom{-}1-1}_{-1\phantom{-}1}$ . When $q=0$ $(\Pi^{0})^{00}_{00}=(\Pi^{0})^{\phantom{-}1-1}_{-1\phantom{-}1} = (\Pi)^{0}$ and this expression simplifies to $\Theta^{-1} \propto 1 - (c_{0}-2c_{2})(\Pi)^{0}$. 

\textit{Results.---} Repulsive spin independent interactions ($c_0$) {\em suppress} both ferromagnetism and pairing in Bose systems.  This should be contrasted to fermions, where due to the opposite sign of the exchange term, repulsive interactions enhance ferromagnetism, giving rise to the Stoner instability \cite{Stoner}, even in the absence of any spin dependent interactions.

 From Eq.~\ref{eq:RPA}, and the fact that $\chi^{0}(0, 0) < 0$ we see that the spin susceptibility only diverges when $c_{2} < 0$ with $|c_{2}| > c_{0}/3$. Similarly, at $q=0$, the pairing susceptibility only diverges when $c_2>0$ with $c_2>c_0/2$.  For weak interactions ($|c_{2}n| \ll k_{B}T$), these instabilities occur near $\mu=0$.  Expanding the susceptibilities for small $\mu$ at $q=0$ gives that, to leading order, the magnetic instability occurs at
 \begin{eqnarray}
t_{\rm mag}&=&  \frac{T_{c}^{\rm mag} - T_{\rm BEC}}{T_{\rm BEC}}=4.84 \left(\frac{1}{3}- \frac{c_{2}}{c_0} \right) n^{1/3} a_{0}\\\nonumber
t_{\rm pair}&=&  \frac{T_{c}^{\rm pair} - T_{\rm BEC}}{T_{\rm BEC}}=6.44 \left( \frac{c_{2}}{c_0} -\frac{1}{2}\right) n^{1/3} a_{0}.
 \end{eqnarray}
Hypothetically, taking $n = 10^{14}$cm$^{-3}$, $a_{0} = 100 a_{B}$, and $|c_2|\sim c_0$, we find $T_{c} - T_{BEC} \sim 10$nK.  The $q=0$ phase diagram is summarized in Fig.~\ref{fig:-1}.

We now explore the role of the quadratic Zeeman effect: 
$q<0$ favors magnetism in the $\pm \hat z$ direction (${\bf F_\parallel}$ -- Ising order -- signalled by diverging $\chi_z$), and pairs in the $m_F=\pm 1$ states (${\bf NP}_\perp:2|\langle \psi_{1}^\dagger \psi_{-1}^\dagger\rangle|>|\langle \psi_0^\dagger \psi_0^\dagger\rangle|$); while $q>0$ favors magnetism in the $\bf x-y$ plane (${\bf F_\perp}$ -- x-y order -- diverging $\chi_\pm$), and $m_F=0$ pairs (${\bf NP}_\parallel: |\langle \psi_0^\dagger \psi_0^\dagger\rangle|>2|\langle \psi_{1}^\dagger \psi_{-1}^\dagger\rangle|$).  Finite $q$ also shifts the BEC transition temperature: the density is given by $n \Lambda_T^3= g_{3/2}(e^{\beta\mu})+2g_{3/2}(e^{\beta(\mu-q)})$, with condensation at $\mu=q$ for $q<0$ and at $\mu=0$ for $q>0$.  For small $q$ one finds
\begin{equation}\label{eq:tcshift}
T_{\rm BEC}^{q\neq0}=T_{\rm BEC}^{q=0}+\xi \sqrt{T_{\rm BEC}^{q=0} q}
\end{equation}
with $\xi=0.3$ for $q<0$ and $\xi=0.6$ for $q>0$.

\begin{figure}
\begin{picture}(100, 90)
\put(-78, 77){\textbf{(a)}}
\put(60, 77){\textbf{(b)}}
\put(60, -3){\includegraphics[scale=0.37]{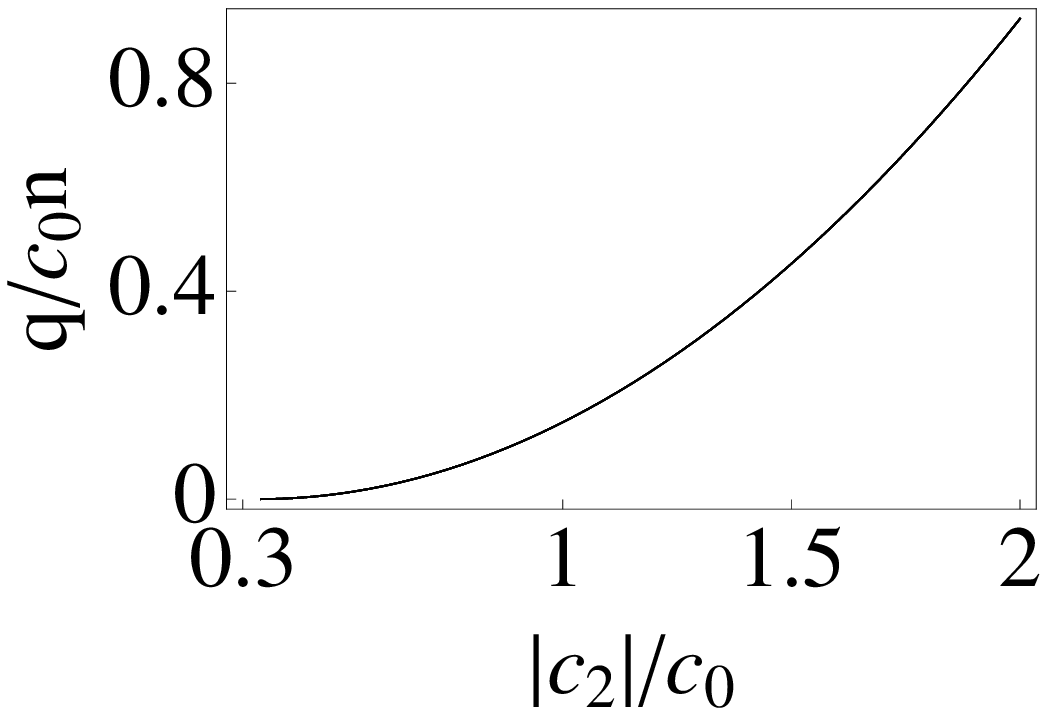}}
\put(-75, -10){\includegraphics[scale=0.45]{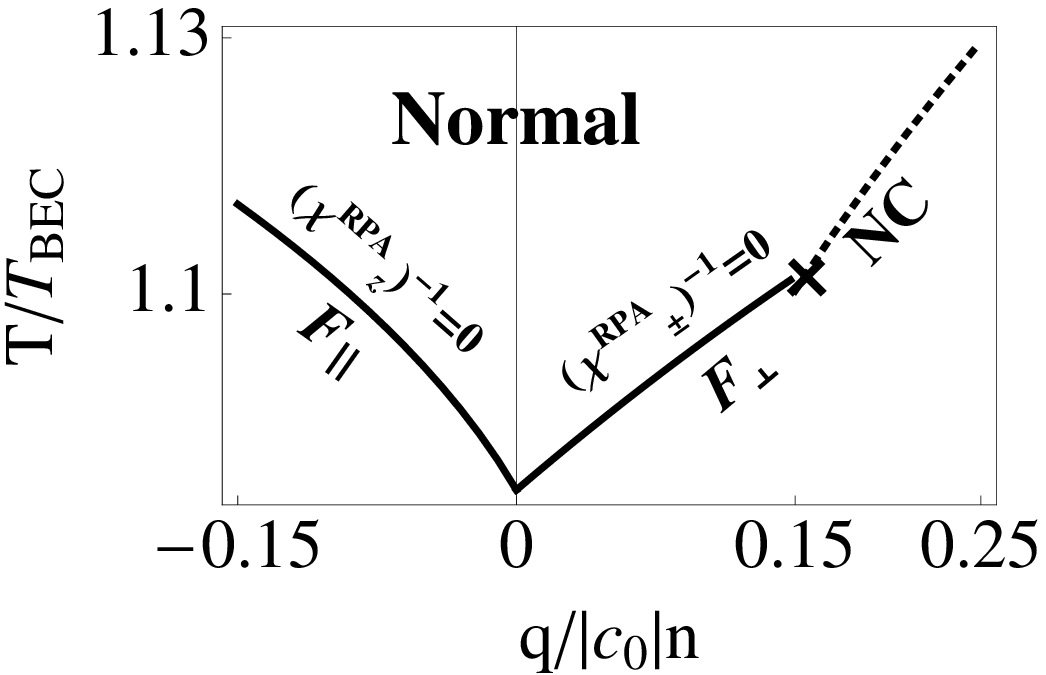}}
\end{picture}

\caption{\label{fig:-2}\textbf{Instability with $\bf c_2 <0$}: (a): Instabilities of the unordered normal state with $c_{2} = -c_{0}$ as a function of $q$. Solid curves give the $T_{c}$ for a non-condensed ferromagnetic gas, normalized to $T_{BEC}|_{q=0}$ (defined in Fig.\ref{fig:-1} caption). At some lower temperature, one expects a transition to a ferromagnetic  condensate (\textbf{FC}). For $q<0$, this $T_{c}$ always exceeds the ideal Bose gas transition temperature. For $q>0$, the ideal gas temperature meets the $T_{c}$ for ferromagnetism at some finite $q$ (marked by $\times$). Beyond this point, the normal state is unstable to forming a polar condensate. (b): Location of  $\times$ as a function of interaction strength.}
\end{figure}

Figure~\ref{fig:-2}(a) illustrates the phase diagram for $c_2<0$, where the only relevant instabilities are ferromagnetism and single particle condensation.  For $q<0$ and $|c_2|>c_0/3$ an Ising ferromagnetic instability always precedes condensation.  For $q>0$ there is a threshold $q$ below which $\bf x-y$ ferromagnetism precedes condensation.  The dependance of this threshold on $c_2$ is shown in Figure~\ref{fig:-2}(b).  For $c_2$ near $-c_0/3$, one finds: $q_{c} = T^{q=0}_{\rm BEC} (10.6 (a_{0}n)^{1/3} \alpha)^{2}$, where $\alpha = 1/3-|a_{2}|/a_{0} $.

\begin{figure}
\begin{picture}(100, 135)
\put(-78, 65){\textbf{(b)}}
\put(65, 60){\textbf{(c)}}
\put(-50, 128){\textbf{(a)}}
\put(-40, 70){\includegraphics[scale=0.38]{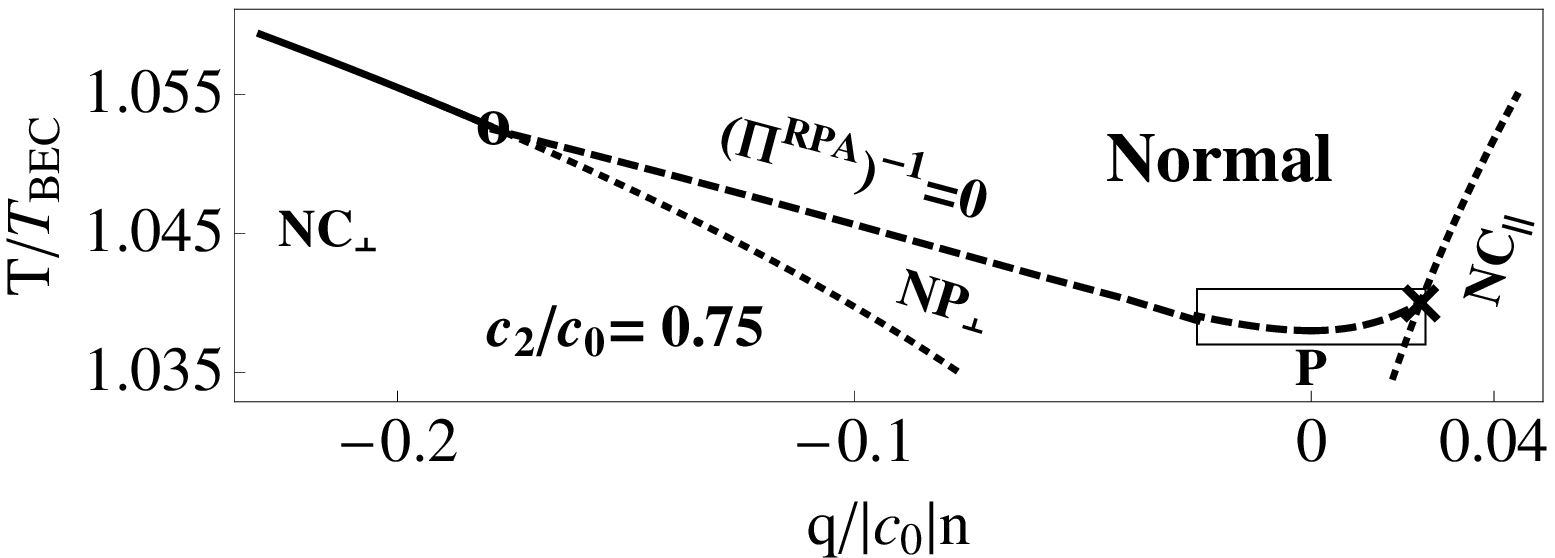}}
\put(-75,-10){\includegraphics[scale=0.39]{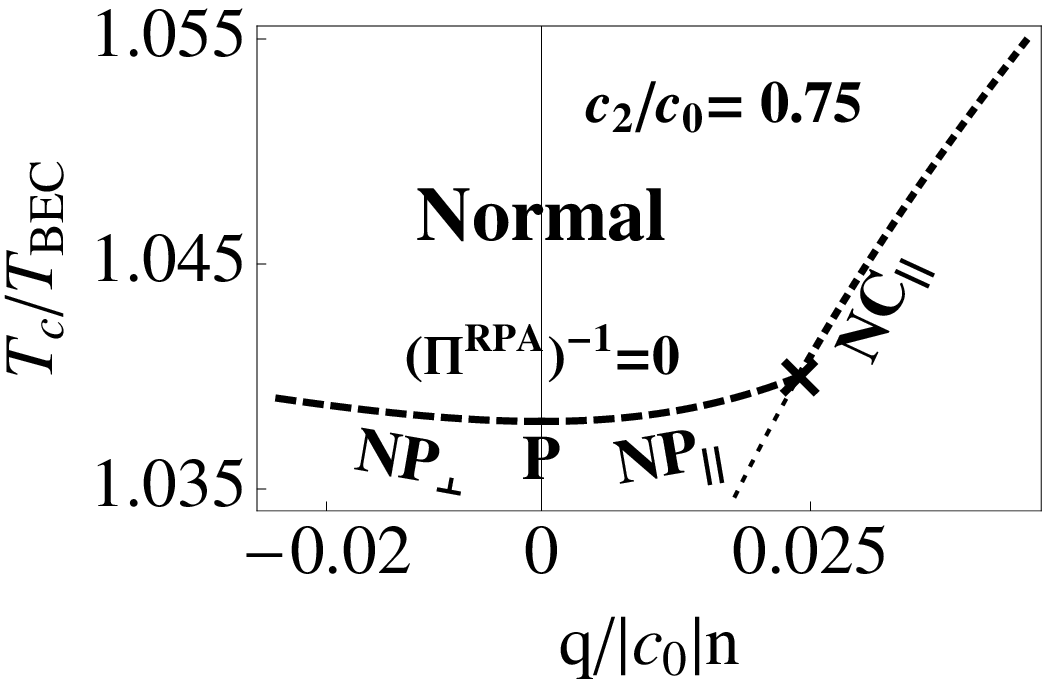}}
\put(66, -10){\includegraphics[scale=0.35]{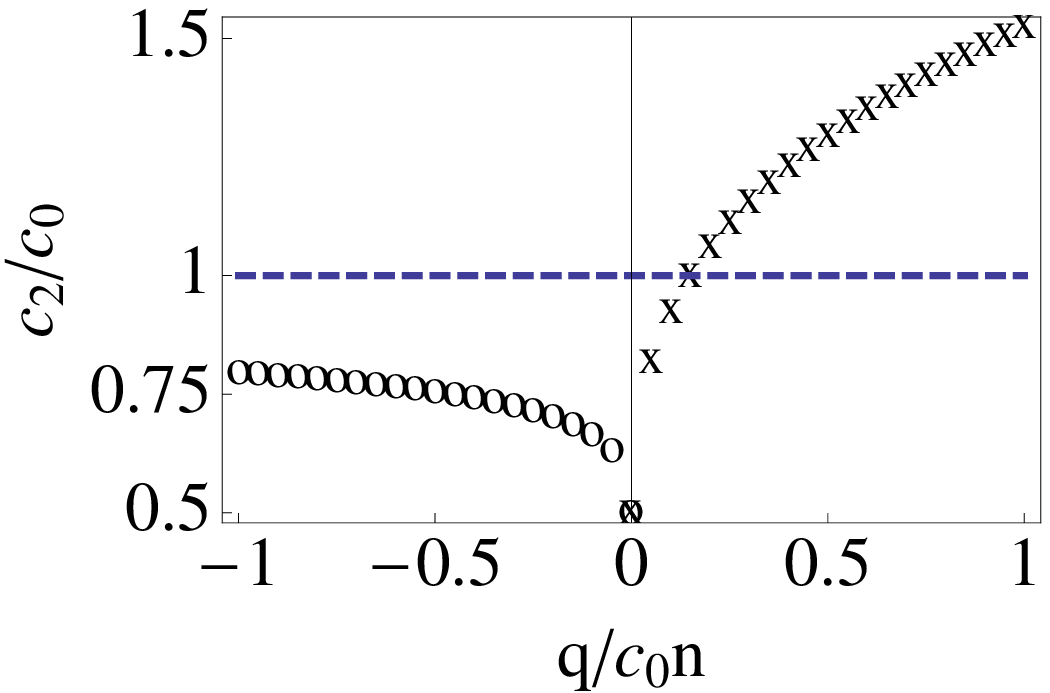}}
\end{picture}

\caption{\label{fig:-3} \textbf{Instabilities with $\bf{c_2 >0}$}: (a): Instabilities of the unordered state with $c_{2} = 0.75c_{0}$. At large negative $q$, the normal state is unstable towards a single-particle condensate ($\textbf{NC}_{\perp}$) with a spinor order parameter $\zeta = \{ e^{i\phi}, 0, 1\})$, with arbitrary $\phi$. For $-|q_{c1}| (\times) < q < |q_{c2}| (o)$ the instability is towards pairing. At $q=0$ the paired phase is a spin singlet with no spin fluctuations. For $q\neq0$, the paired phase has spin fluctuations in the $\bf{x-y}$ plane. For large $q>0$, the normal state is unstable towards a polar condensate with spinor $\zeta = \{0, 1, 0\}$. Dotted lines are the ideal gas condensation $T_{c}$ as a function of $q$. (b): Details of dashed rectangle in (a), showing instabilities towards pairing. (c): Critical values of $q$ at which pair instability gives way to single particle instability (\textit{i.e.} locations of $\times$ and $o$) for different interaction strengths.}
\end{figure}

Figure~\ref{fig:-3}(a) illustrates the phase diagram for $c_2>0$, where the only relevant instabilities are pairing and single particle condensation.  Finite $q$ enhances single particle condensation, and for a given $q$, there is a threshold value of $c_2$ required to find a pairing instability.  For $q>0$ this threshold becomes arbitrarily large as $q\to\infty$, but for $q<0$ one always has a pairing transition if $c_2>c_0$. Setting $\mu = 0$, and taking the limit $(\Pi^{0})^{\phantom{-}1-1}_{-1\phantom{-}1} \rightarrow \infty$ for $q < 0$ and $(\Pi^{0})^{00}_{00} \rightarrow \infty$ for $q>0$, we calculate these threshold values (Fig.~\ref{fig:-3}(b)): 
\begin{equation}
q_{c}  = 12.74T^{q=0}_{BEC}(n^{1/3}a_{0})^2(1-2x)(1+x)\left\{\begin{array}{lr}
1, & q > 0\\
\frac{1}{1-x}, & q < 0
\end{array}\right\}\end{equation}
where $x = c_{2}/c_{0}$. 

Experimentally, the states discussed in Fig.\ref{fig:-3} may be somewhat distinguished by the fact that both the condensed phase $\textbf{NC}_{\perp}$ for large $q<0$, and the paired phase $\textbf{NP}_{\perp}$, have $n_{s} = n_{1}+n_{-1} - 2n_{0} >0$. In the singlet pair, this quantity is identically zero.  Studying momentum distributions can distinguish between the single particle and paired condensates. 

\textit{Finite Size effects.---} 
Similar to \cite{erich1}, we can estimate the role of finite size effects by looking at the instabilities at finite $k=2\pi/L$, where $L$ is the size of the cloud.  These finite size effects are crucial in the scalar gas with attractive interactions, where there is no Bose-Einstein condensation in the thermodynamic limit.  Here we find only small corrections to location of the threshold value of $|c_2|/c_0$. To leading order in $k\Lambda_{T} \ll 1$, the threshold for ferromagnetism is
$a_{2}/a_{0} = -\frac{1}{3} - \frac{k \Lambda^{2}_{T_{BEC}}}{12\pi^{2} a_{0}} $, while the  threshold for pairing is
$a_{2}/a_{0}=  \frac{1}{2} + \frac{k\Lambda^{2}_{T}}{8\pi^{2}a_{0}}$.  These shifts are no greater than $10\%$ for a system of size $L\sim 100\mu$m.

\textit{Conclusions.---} The presence of competing magnetic and off-diagonal long range orders in a spin $1$ gas produces an extremely interesting phase diagram with ferromagnetic, nematic and paired phases.  Using the RPA-X, we have quantitatively studied the instabilities of the normal state, identifying the temperatures and interaction strengths at which the disordered normal state becomes unstable to a symmetry broken phase. We find that the finite temperature phase diagram is featureless unless the interaction strengths governing spin ($c_{2}$) and charge physics ($c_{0}$) are comparable in magnitude. This is due to the exchange enhancement of identical particle scattering. 

A number of probes can be used to distinguish the nematic phases and detect pairing. These include optical birefringence \cite{carusotto}, momentum distributions via time-of-flight, noise correlations \cite{greiner}, and the nature of vortices. 

Finally, we remark that the key bottleneck to realizing this interesting phase diagram is that of finding atoms with spin independent and spin-dependent interactions of comparable magnitude. The search for such atoms is an active area of research, and we hope that our work motivates this effort. 

\textit{Acknowledgements.---} We thank Stefan K. Baur, Joel E. Moore and Xiaoling Cui, Ari Turner and Mukund Vengalattore for useful conversations. This material is based upon work supported by the National Science Foundation through grant No. PHY-0758104.

\section{SUPPLEMENTARY MATERIALS FOR ``Instability of a spin-$1$ Bose gas to ferromagnetism and pairing"}
\numberwithin{equation}{section}
\renewcommand{\theequation}{A-\arabic{equation}}

\textit{Analytic structure of non-interacting response functions.---}
We develop the analytic structure of $\chi^{0\beta\alpha}_{\alpha\beta}$ defined in Eq.\ref{eq:4} of the main text as:
\begin{equation}\label{eq:supp1}
(\chi^{0})^{\beta\alpha}_{\alpha\beta}(p, \omega) = \int \frac{d^{3}\textbf{k}}{(2\pi)^{3}}\frac{n(\epsilon_{k, \alpha}) - n(\epsilon_{k+p, \beta})}{\omega - (\epsilon_{k+p \beta} - \epsilon_{k\alpha})}
\end{equation}

For $q=0$ this simplifies to a constant $\chi(p, \omega) = \int \frac{d^{3}\textbf{k}}{(2\pi)^{3}}\frac{n(\epsilon_{k}) - n(\epsilon_{k+p})}{\omega - (\epsilon_{k+p} - \epsilon_{k})}$, whose structure has been thoroughly explored in \cite{erichstab}. The result  to linear order in $p\Lambda_T$ is :
\begin{eqnarray}\label{eq:supp2}
\chi(\textbf{p}, 0) = -\frac{m}{2\pi\lambda_{T}}\left(g_{1/2}(e^{\beta\mu}) - \sqrt{|\pi/(\beta\mu)|} + \right. \\\nonumber \left. \frac{i\pi}{p \Lambda_{T}} \log\left(\frac{\sqrt{\beta\mu} - \frac{p\Lambda_{T}}{4\sqrt{\pi}}}{\sqrt{\beta\mu}+\frac{p\Lambda_{T}}{4\sqrt{\pi}}}\right)\right)
\end{eqnarray}

where $g_{\nu}(z)$ is the polylogarithm function. For $q\neq 0$, first note that $\chi^{\beta\alpha}_{\alpha\beta}(0, 0) = \chi(0,0)|_{\mu \rightarrow \mu -q}$ for $\{\alpha,\beta\} = \{1, -1\}$. The transverse spin response however given by $(\chi^{0})^{10}_{01}(p, \omega)$ requires some work. One proceeds as follows: Integrating out the angular variables one finds
\begin{eqnarray}\label{eq:supp3}
(\chi^{0})^{10}_{01\textbf{p}}(\omega)= \frac{-m}{\pi\lambda_{T}^{2}p}\int^{\infty}_{0}d\tilde k n(\epsilon_{\tilde k1})\log\left(\frac{z_{-}+\tilde k}{z_{-}+\tilde k}\right) \\\nonumber +~n(\epsilon_{\tilde k0})\log\left(\frac{z_{+}-\tilde k}{z_{+}+\tilde k}\right)
\end{eqnarray}
where $z_{\pm} = \frac{\omega + q}{2\sqrt{\epsilon_{p}k_{B}T}} \pm \frac{p\Lambda_{T}}{4\sqrt{\pi}}$ and $\tilde k = k\Lambda_{T}/4\sqrt{\pi}$. We use $\tilde k$ as an expansion parameter. 

Rewriting the logarithm as an integral we get:
\begin{equation}\label{eq:supp4}
(\chi^{0})^{10}_{01}(p, \omega) = -\frac{m}{2\pi\lambda_{T}^{2}p}\left(\int^{z_{-}}_{\infty}dx I_{1}(k)- \int^{z_{+}}_{\infty}dxI_{0}(k)\right)
\end{equation}
where $I_{1/0}(k) = \int^{\infty}_{-\infty}dk \frac{1}{k-x}\frac{2k }{e^{k^{2} + \beta\lambda_{1/0}} -1}$, where $\lambda_{1} = -\beta(\mu - q)$ and $\lambda_{0} = -\beta\mu$.
The analytic structure of the integral $I$ has been extensively developed by Sz\'epfalusy and Kondor \cite{sk}, who show that the integral can be written as an asymptotic series for long wave-lengths. Retaining only the lowest order terms we find that the static response yields:
\begin{equation}\label{eq:supp5}
(\chi^{0})^{10}_{01}(0, 0) = \frac{m}{4\pi\lambda_{T}}\frac{k_{B}T}{q}(g_{3/2}(e^{\beta(\mu-q)}) - g_{3/2}(e^{\beta\mu}))
\end{equation}

The non-interacting pair response is defined as: 
\begin{equation}\label{eq:supp6}
(\Pi^{0})^{\alpha\beta}_{\beta\alpha}(\textbf{p}, \omega) = \int \frac{d^{3}\textbf{k}}{(2\pi)^{3}}\frac{n(\epsilon_{\textbf{k}, \alpha}) + n(\epsilon_{\textbf{k}+\textbf{p}, \beta})}{\omega - (\epsilon_{\textbf{k}+\textbf{p} \beta} + \epsilon_{\textbf{k}\alpha})}
\end{equation}

Once again, the pair susceptibility for the scalar gas has been considered in \cite{erich1}. The result to linear order in $k\Lambda_{T}$ is :
\begin{eqnarray}\label{eq:supp7}
\Pi^{0}(\textbf{p}, 0) = -\frac{m}{\pi\lambda_{T}}\left( g_{1/2}(e^{\beta\mu}) - \sqrt{|\pi/(\beta\mu)|} \right. \\\nonumber + \left. \frac{2i\pi}{p \Lambda_{T}} \log\left(\frac{(1+i)\frac{p\Lambda_{T}}{4\sqrt{\pi}}-\sqrt{\beta\mu}}{(1-i)\frac{p\Lambda_{T}}{4\sqrt{\pi}}-\sqrt{\beta\mu}}\right)+ \frac{p \Lambda_{T}}{8} \right)\end{eqnarray}

It is easy to show that $(\Pi^{0})^{\phantom{-}1-1}_{-1\phantom{-}1}(0, 0) = \Pi^{0}(0, 0)|_{\mu \rightarrow \mu-q}$ and $(\Pi^{0})^{00}_{00} = \Pi^{0}$. 

\textit{Spin response in the RPA .---}
The spin response in the RPA is given by solving for the polarization tensor defined in Eq.~\ref{eq:2} using Eq.~\ref{eq:10}. The interaction matrix $V$ encompasses all direct and exchange diagrams and takes the form:
\begin{widetext}
\begin{eqnarray}
V = \left(\begin{array}{lllcccrrr}
2(c_{0} + c_{2}) & 0 & 0&0&c_{0}+c_{2}&0&0&0&c_{0}-c_{2} \\
0&0&0&c_{0}+c_{2}&0&0&0&2 c_{2}&0 \\
0&0&0&0&0&0&c_0 - c_2&0&0 \\
0&c_0+c_2&0&0&0&2c_2&0&0&0 \\
c_0+c_2&0&0&0&2 c_{0}&0&0&0&c_0 +c_2\\
0&0&0&2c_2&0&0&0&c_0+c_2&0\\
0&0&c_0-c_2&0&0&0&0&0&0\\
0&2c_2&0&0&0&c_0+c_2&0&0&0\\
c_0-c_2&0&0&0&c_0+c_2&0&0&0&2(c_0 + c_2)
\end{array}\right)
\end{eqnarray}
\end{widetext}

From the full polarization tensor, one extracts the longituginal and transverse spin susceptibility on which our calculations are based.

We now turn to the details of the pair response calculation.

\textit{Pairing response in the RPA.---}
The RPA response is given by Eq.~\ref{eq:10} where $V$ is a symmetric $9 \times 9$ matrix. However since pairing only occurs in the $S_{z} =0$ channel, it suffices to consider the following subsystem
\begin{eqnarray}\label{eq:supp8}
\left(\begin{array}{lcr} 
\Pi^{\phantom{-}1-1}_{-1\phantom{-}1} & \Pi^{1-1}_{0\phantom{-}0} & \Pi^{1-1}_{1-1}\\
\Pi^{0\phantom{-}0}_{1-1} & \Pi^{00}_{00} & \Pi^{\phantom{-}00}_{-11}\\
\Pi^{-11}_{-11} & \Pi^{-11}_{\phantom{-}00} & \Pi^{-1\phantom{-}1}_{\phantom{-}1-1}
\end{array}\right)^{RPA}\end{eqnarray}
which is related to the non-interacting response $\left(\begin{array}{lcr} \Pi^{\phantom{-}1-1}_{-1\phantom{-}1} & 0 &0\\
0 & \Pi^{00}_{00} & 0\\
0 & 0 & \Pi^{-1\phantom{-}1}_{\phantom{-}1-1}
\end{array}\right) 
$ via Eq.~\ref{eq:10} and the $3 \times 3$ interaction matrix $ V= \left(\begin{array}{lcr} c_0 - c_2 & c_2 &0\\
c_2 & c_0 & c_2\\
0 & c_{2} & c_0 - c_2
\end{array}\right)$.

Note that $\Pi^{\phantom{-}1-1}_{-1\phantom{-}1} = \Pi^{-1\phantom{-}1}_{\phantom{-}1-1}$ and the physical response of the system to adding a $\pm1$ pair is given by $\Pi^{\phantom{-}1-1}_{-1\phantom{-}1}  +\Pi^{-1\phantom{-}1}_{\phantom{-}1-1}$. The $3 \times 3$ system can be further reduced to a $2 \times 2$ system of equations which then yields the singlet RPA response. 

Two limiting cases are worth considering. The first is at $\mu = q$ for $q<0$ when the non-interacting BEC transition occurs for the $\pm1$ atoms.  At these values the non-interacting response function $\Pi^{\phantom{-}1-1}_{-1\phantom{-}1}$ diverges, and:
\begin{equation}\label{eq:supp9}
\Theta^{-1} \propto- c_{0} + c_{2} +  (c_0 - 2 c_2)(c_{0} + c_{2})\Pi^{00}_{00}
\end{equation}
The second is at $\mu = 0$ which corresponds to the BEC transition for the $0$ atoms. At this point non-interacting response function $\Pi^{00}_{00}$ diverges and:  \begin{equation}\label{eq:supp10}
\Theta^{-1} \propto- c_{0} +  (c_0 - 2 c_2)(c_{0} + c_{2})\Pi^{\phantom{-}1-1}_{-1\phantom{-}1}
\end{equation}

Setting the R.H.S. of Eqs.~(\ref{eq:supp9}, \ref{eq:supp10}) to zero, using Eq.\ref{eq:tcshift} along with the functional forms of the response functions, yields the threshold value of $c_2$ at which paired states result for any $q$.

\end{document}